\documentclass{aa}

\input psfig.sty
\usepackage{psfig}

\newcommand{\be}{\begin{equation}}
\newcommand{\ee}{\end{equation}}
\newcommand{\bea}{\begin{eqnarray}}
\newcommand{\eea}{\end{eqnarray}}

\newcommand{\ergpers}{\hbox{\rm erg$\,$s$^{-1}$}}
\newcommand{\kms}{\hbox{\rm km$\,$s$^{-1}$}}

\def\aap{{A\&A}}
\def\aapl{{A\&A} (Letters) }
\def\aaps{{A\&AS}}

\def\apj{{ApJ}}

\def\apjs{{ApJS}}

\def\mnras{{MNRAS}}

\def\ssr{{Space Sci. Rev. }}

\def\la{\mathrel{\hbox{\rlap{\hbox{\lower4pt\hbox{$\sim$}}}\hbox{$<$}}}}
\def\ga{\mathrel{\hbox{\rlap{\hbox{\lower4pt\hbox{$\sim$}}}\hbox{$>$}}}}
\def\ea{{et al.} }

\newcommand{\gired}{$\chi^2_{\rm red}$}

\newcommand {\Cvi}{C\,{\sc vi} }

\newcommand {\Nvi}{N\,{\sc vi} }
\newcommand {\Nvii}{N\,{\sc vii} }

\newcommand {\Ovii}{O\,{\sc vii} }
\newcommand {\Oviii}{O\,{\sc viii} }

\newcommand {\Sixiii}{Si\,{\sc xiii} }

\newcommand {\Neix}{Ne\,{\sc ix} }
\newcommand {\Nex}{Ne\,{\sc x} }

\newcommand {\Mgxi}{Mg\,{\sc xi} }

\newcommand {\Sxv}{S\,{\sc xv} }

\newcommand {\Fexvii}{Fe\,{\sc xvii} }

\def\la{\mathrel{\hbox{\rlap{\hbox{\lower4pt\hbox{$\sim$}}}\hbox{$<$}}}}
\def\ga{\mathrel{\hbox{\rlap{\hbox{\lower4pt\hbox{$\sim$}}}\hbox{$>$}}}}

\def\ion#1#2{#1\,{\sc #2}}

\begin{document}

\title{XMM-Newton observations of $\zeta$ Orionis (O9.7 Ib):\\ A Collisional Ionization Equilibrium model}

\author{
A.J.J.~Raassen\inst{1,2}
\and K.A.~van der Hucht\inst{1,2} 
\and N.A.~Miller\inst{3}
\and J.P.~Cassinelli\inst{4} 
}

\institute{ 
SRON Netherlands Institute for Space Research, Sorbonnelaan 2,                   
                NL 3584 CA Utrecht, The Netherlands
\and
     Astronomical Institute "Anton Pannekoek", Kruislaan 403, NL 1098 SJ Amsterdam,
                The Netherlands
\and 
     Dept. of Physics \& Astronomy, University of Wisconsin-Eau Claire, 
     105 Garfield Avenue, Eau Claire WI 54702, USA 
\and
     Dept. of Astronomy, University of Wisconsin at Madison, 6251 Sterling Hall,
     North Charter Str., Madison, WI 53706, USA
}

\offprints{A.J.J.~Raassen \email{a.j.j.raassen@sron.nl}}
\date{Received / Accepted}
\date{ {} / \today}

\abstract{We present {\em XMM-Newton} observations of the O supergiant $\zeta$~Orionis (O9.7 Ib).
The spectra are rich in emission lines over a wide range of ionization stages.
The {\sc rgs} spectra show for the first time lines of low ion stages such
as \ion{C}{\sc vi}, \ion{N}{\sc vi}, \ion{N}{\sc vii}, and \ion{O}{\sc vii}. 
The line profiles are symmetric and rather broad (FWHM $\approx$ 1500 km s$^{-1}$) and 
show only a slight blue shift. With the {\sc XMM-epic} spectrometer several high ions are detected in this star for the first time
including \ion{Ar}{\sc xvii} and \ion{S}{\sc xv}. \\
Simultaneous multi-temperature fits and DEM-modeling were applied to the {\sc rgs} and {\sc epic} spectra to obtain 
emission measures, elemental abundances and plasma temperatures. The calculations show
temperatures in the range $\approx$0.07~-~0.6~keV.
According to the derived models the intrinsic source X-ray luminosity at a distance of 251 pc 
is $L_{\rm x}$\,=\,1.37(.03)\,$\times$\,10$^{32}$\, \ergpers\, in the energy range 0.3~-~10 keV.
In the best multi-temperature model fit, the abundances of C, N, O, and Fe are near 
their solar values, while the abundances of Ne, Mg, and Si appear somewhat enhanced.\\
The sensitivity of the He-like forbidden and intercombination lines to $\zeta$~Ori's 
strong radiation field is used to derive the
radial distances at which lines are formed. Values of 34~$R_*$ for \ion{N}{\sc vi}, 
12.5~$R_*$ for \ion{O}{\sc vii}, 4.8~$R_*$ for \ion{Ne}{\sc ix}, and 3.9~$R_*$ for  \ion{Mg}{\sc xi}
are obtained. \\  
}
\titlerunning{XMM observations of $\zeta$~Ori}
\authorrunning{A.J.J.Raassen et al.}
\maketitle
%\markboth{}{}

\begin{table*}[ht]
\caption{Observation log of the data of $\zeta$~Ori.}
\begin{center}
\begin{tabular}{|l@{\ }|l@{\ }|l@{\ }|l@{\ }|l@{\ }|r@{\ }|}
\hline
Instr.           &  Filter&Mode&Date-obs-start&Date-obs-end&Duration(s)\\
\hline
MOS1                 &  Thick &Small Window&2002-09-15T13:12:32&2002-09-16T00:51:00&41728\\
MOS2                 &  Thick &Timing      &2002-09-15T13:12:28&2002-09-16T00:46:41&41473\\
pn                   &  Thick &Full Frame  &2002-09-15T14:04:08&2002-09-16T00:51:20&38362\\
RGS1                 &  None  &Spec+Q      &2002-09-15T13:11:17&2002-09-16T00:52:38&41979\\
RGS2                 &  None  &Spec+Q      &2002-09-15T13:11:17&2002-09-16T00:52:38&41979\\
\hline
\end{tabular}
\end{center}

\end{table*}

\begin{figure*}
\hbox{\psfig{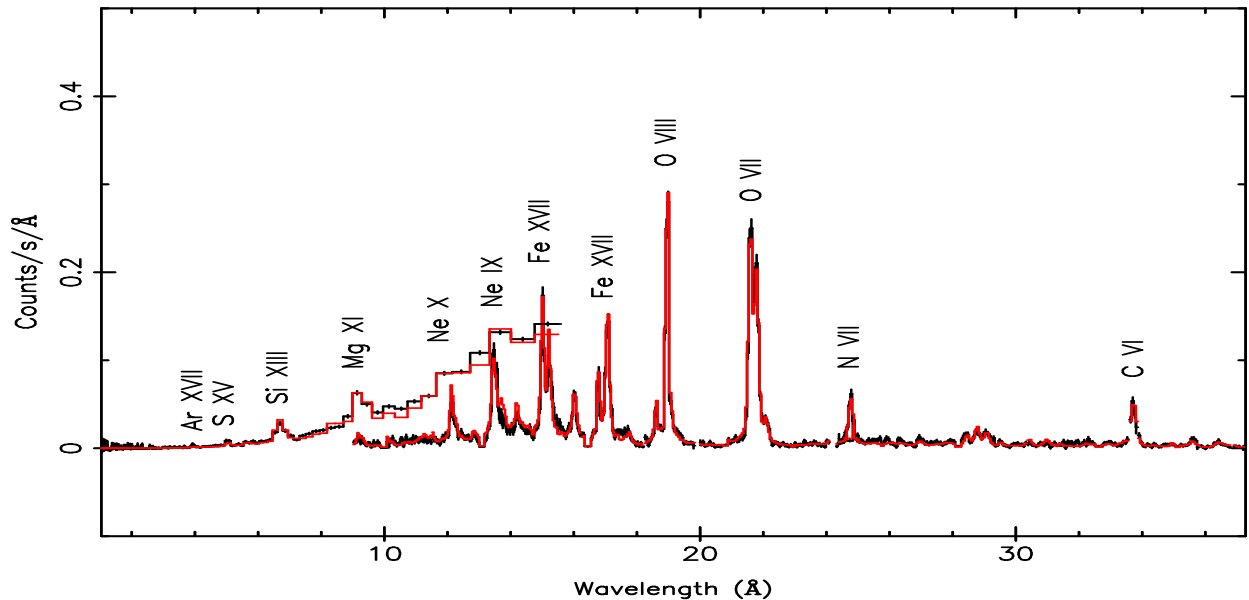}}
      \caption[]{Background-subtracted {\sc epic-mos} and first-order {\sc rgs} spectra
      of $\zeta$~Ori. 
      Error bars indicate 1$\sigma$ statistical errors including the effect of background.
      The grey (red in the electronic version) curve shows the best-fit model. A number of prominent lines are labeled with the 
      emitting ions.
              }
         \label{spectrum}
   \end{figure*}

\section{Introduction\label{sec_intro}}
The object $\zeta$~Orionis is a supergiant of spectral type O9.7~Ib.
Its X-ray spectrum has previously been observed
with various satellites such as {\em Einstein} (Cassinelli \& Swank
1983), {\em ROSAT} (Bergh\"{o}fer et al. 1997), 
and {\em ASCA} (Kitamoto et al. 2000). 
The {\em Einstein} observations were particularly interesting: the Solid-State Spectrometer
observations of $\zeta$~Ori were the first to show hot star X-ray emission lines,
those of \Sixiii and \Sxv (Cassinelli \& Swank 1983).  

From {\em ROSAT} observations Bergh\"{o}fer et al. (1996) found that the X-ray emitting 
plasma could be characterized with an overall $kT$ of 0.22~keV and that 
the X-ray to Bolometric luminosty ratio
of log $L_{\rm x}/L_{\rm bol}$ = -6.74 is typical 
for O stars and early B supergiants.
Based on {\em Chandra}-{\sc hetg} data Waldron \& Cassinelli (2001) concluded that most of 
the emission is at energies below 1 keV.

It is generally believed that most of these X-rays received from O stars originate 
in shocks distributed throughout the wind, and that these shocks are created by 
instabilities in the line-driving mechanism which drives the overall 
outflow (e.g., Lucy \& Solomon 1970, Lucy 1982, Owocki et al. 1988, Dessart \& Owocki 2005). 
To explore the relation between the X-ray emission from $\zeta$~Ori and the properties of 
the shocks in its wind line profiles from high-resolution {\em Chandra} observations were 
used by Cohen et al.
(2006), Owocki \& Cohen (2006), and Oskinova et al. (2006).

Waldron \& Cassinelli (2001) also analyzed the ratios of the forbidden($f$), the intercombination($i$) and the
resonance($r$) emission lines in the He-like ions of \ion{Si}{\sc xiii},
\ion{Mg}{\sc xi}, \ion{Ne}{\sc ix}, and \ion{O}{\sc vii}.
The lines were found to be produced at a wide range of radii.
The specific radius for each He-like ion was
found to correspond to X-ray optical depth unity to the continuum opacity at
each line wavelength. Their most surprising result was that the high ion
stages, such as \ion{Si}{\sc xiii}, appear to form very close to the star. At
these close distances the shock velocity jump was found to be larger than
could be explained by the speed that the wind should have at these small
radii.

Recently Pollock (2007a) developed a new paradigm for O-star X-ray emission, which
states that charge exchange, ionization, and excitation are likely to be produced by protons far out in the wind. However, this approach
is not yet capable of describing the spectrum and line fluxes in a quantitative way, and so it will not be discussed
further in this paper.

Variability has been found in both $\zeta$~Ori's H$\alpha$ profile
(Ebbets 1982; Kaper et al. 1997) and its UV P-Cygni profiles (Snow 1977; Kaper
et al. 1996; Kaper et al. 1999). However, the X-ray light curve has provided no evidence for 
variability over a long time range (Bergh\"ofer \& Schmitt 1994).

In many of the recent papers the authors have focussed on line profiles and individual line features. This is
especially true for spectra obtained by means of the high resolution instruments aboard {\em Chandra} ({\sc hetg} and {\sc meg}). 
In contrast {\sc rgs} aboard {\em XMM-Newton} provides us with new information regarding the longer wavelength
part of the spectrum where ion stages such as \ion{C}{\sc vi}, \ion{N}{\sc vi}, and \ion{N}{\sc vii} can be observed. In
addition, the higher collecting area of {\em XMM-Newton} allows us to obtain a better
estimate of the distribution of emission measure versus temperature.
In the present paper the spectra and line-fluxes of $\zeta$~Ori, observed with {\em XMM}-{\sc epic} and -{\sc rgs}, are described for the first
time by means of a multi-temperature fit and DEM-modeling to the total spectrum.
The approach we choose to take is to use emissivity results from optically thin
Coronal Ionization Equilibrium (CIE) plasma calculations.

Apart from this broad description of the total spectrum, individual line fluxes and
line ratios are analyzed in this paper. 
Ratios of the $fir$ lines of He-like ions are used to determine the typical distances
between the X-ray emission regions and the stellar photosphere.
The wavelength sensitivity of 
{\em XMM-Newton} allows us to check the results of Waldron \& Cassinelli (2001) 
for \ion{Mg}{\sc xi}, \ion{Ne}{\sc ix}, and \ion{O}{\sc vii}
(with somewhat more precise radial ranges), while extending the analysis to include the \ion{N}{\sc vi} lines.
The results for the lines of \ion{N}{\sc vi} are particularly interesting because 
this is the first time that this line ratio has been interpreted in this 
manner for this star.

$\zeta$~Ori was observed on 2002 September 9 by means of {\em XMM-Newton}.
A log of these observations is shown in Table~1.
The spectral analysis by means of multi-temperature fitting and DEM-modeling are described in Sect.~2. 
The analysis of the
line profiles and the individual line fluxes, including several that have not been studied earlier, 
is given in Sect.~3. The results are discussed in Sect.~4.
Parameters of $\zeta$~Ori are given in Table~A.1 in the appendix.

\section{Spectral analysis}

\subsection{Multi-temperature fitting}
   
\begin{table}[t!]
\caption{Multi-temperature fitting for combined {\sc epic-pn}, {\sc epic-mos1} \& {\sc rgs} spectra of $\zeta$~Ori$^a$.} 
\begin{center}
\begin{tabular}{   l@{\ }|l@{\ }  l@{\ }  }
\hline \hline
comp. i                           & value	   & \\[+0.3mm]	  
\hline
$N_{\rm H}$$^b$                   & 5.0 	   & \\[+0.3mm]	  
 	    \\[+0.3mm]   
$kT_{\rm 1}$$^c$                  &0.551$\pm$0.013 & \\[+0.3mm]
$EM_{\rm 1}$$^d$                  &1.57$\pm$0.18   & \\[+0.3mm]
$L_{\rm x1}$$^e$                  &4.18 	   & \\[+0.3mm]  
$L_{\rm x1}$$^f$                  &3.8 	   & \\[+0.3mm]
$v_{\rm mic1}$$^g$                &1080$\pm$135 	   &  \\[+0.3mm]
 	    \\[+0.3mm]
$kT_{\rm 2}$$^c$                  &0.201$\pm$0.004  & \\[+0.3mm]
$EM_{\rm 2}$$^d$                  &3.76$\pm$0.31    & \\[+0.3mm]
$L_{\rm x2}$$^e$                  &6.71 	    & \\[+0.3mm]  
$L_{\rm x2}$$^f$                  &20.6 	  &  \\[+0.3mm]
$v_{\rm mic2}$$^g$                &950$\pm$70	    & \\[+0.3mm]
 	    \\[+0.3mm]
$kT_{\rm 3}$$^c$                  &0.073$\pm$0.006  & \\[+0.3mm]
$EM_{\rm 3}$$^d$                  &10.1$\pm$2.7     & \\[+0.3mm]
$L_{\rm x3}$$^e$                  &2.8 	   &  \\[+0.3mm]  
$L_{\rm x3}$$^f$                  &111.0 	   & \\[+0.3mm]
$v_{\rm mic3}$$^g$                &1240$\pm$230    &  \\[+0.3mm]
 	    \\[+0.3mm]

$\sum_{\rm i} EM_{\rm i}$$^d$     &15.4$\pm$2.2    &  \\[+0.3mm]
$\sum_{\rm i} L_{\rm xi}$$^e$     &13.7$\pm$0.3     & \\[+0.3mm] 
$\sum_{\rm i} L_{\rm xi}$$^f$     &135$\pm$4     & \\[+0.3mm] 
$z_{\rm red}$$^h$                 &-3~10$^{-4}$     & \\[+0.3mm] 
\hline
                                  &Abundance$^i$  & \\[+0.3mm]                        
                                  & $A_i$         & $A_i/A_{\rm O}$\\[+0.3mm]     
C                                 &1.02$\pm$0.11  &1.20$\pm$0.05\\[+0.3mm]
N                                 &1.03$\pm$0.09  &1.21$\pm$0.05\\[+0.3mm]
O                                 &0.85$\pm$0.07  &1\\[+0.3mm]
Ne                                &1.33$\pm$0.15  &1.57$\pm$0.07\\[+0.3mm]
Mg                                &1.66$\pm$0.16  &1.96$\pm$0.08\\[+0.3mm]
Si                                &1.30$\pm$0.14  &1.54$\pm$0.05\\[+0.3mm]
Fe                                &0.99$\pm$0.09  &1.15$\pm$0.06\\[+0.3mm] 
\hline
\gired                            &1246/1064=1.17 \\[+0.3mm]
\hline\hline
\end{tabular}
\end{center}
\begin{flushleft}
{
\begin{description}
\item Notes:
\item $a$: {\sl HIPPARCOS} distance $d$ = 0.25 kpc is used.
\item $b$: Column density (10$^{20}$~cm$^{-2}$) derived for standard absorption model (with solar abundances) from 
  Morrison \& McCammon (1983) for all components.
\item $c$:  Temperature in keV. 
\item $d$:  Emission measure in 10$^{54}~$cm$^{-3}$. 
\item $e$:  X-ray luminosity of the instrumental {\em XMM-Newton} band (0.3--10~keV) in 10$^{31}$ \ergpers.
\item $f$:  X-ray luminosity of the total emission range, 0.03--0.9~keV (see Fig.~\ref{DEM}), in 10$^{31}$ \ergpers.
\item $g$:  Micro-turbulent velocity parameter used to characterize line broadening (in \kms).
\item $h$:  Artificial redshift to match the observed wavelengths to the theoretical ones.
\item $i$: Abundance relative to solar photospheric number abundance (Anders \& Grevesse 1989 for all ions 
           except Fe, where Grevesse \& Sauval 1998, 1999 was used). 
\end{description}
}
\end{flushleft}
\end{table}

We have determined the thermal structure and the elemental composition of
$\zeta$~Ori's X-ray emitting plasma by simultaneously fitting multi-temperature models to 
the {\sc rgs}, {\sc epic-mos1}, and {\sc pn} spectra.
The fit was performed using our spectral analysis program {\sc
spex} (Kaastra et al.~1996a) in combination with the {\sc mekal}
(Mewe-Kaastra-Liedahl) code, as developed by Mewe \ea (1985, 1995)(see Fig.~\ref{spectrum}). 
{\sc mekal} calculates a continuum and models more than
5400 spectral lines. It is available on the
{\sc www}\footnote{\textrm{http://www.sron.nl/divisions/hea/spex/}}.
The applied model is a Collisional Ionization Equilibrium (CIE) model for optically thin plasma. The
ionization equilibrium is based on calculations by Arnaud \& Rothenflug (1985) and Arnaud \& Raymond (1992). 

\subsubsection{Temperature and Emission measure}
In the multi-temperature calculations we used initially five temperatures. 
Out of the five fitted
temperature components three turn out to be significant. 
These three temperature
components were coupled to one $N_{\rm H}$ absorption column
density and one set of abundances, which were free to vary.  The column density, temperatures,  
emission measures $EM=\int n_en_{\rm H}dV$, X-ray~luminosities, elemental abundances, and micro-turbulent velocity ($v_{\rm mic}$)
are given in Table~2, together with the statistical 1\,$\sigma$ uncertainties. Three temperatures 
at $T$=0.073~keV, $T$=0.201~keV and $T$=0.551~keV were determined. 
Interpreting the results in terms of a wind with
embedded shocks and using 
$kT = 1.2\ [v (1000\ {\rm km~s}^{-1})]^2$ keV, we find that the $kT$ range from
0.073~keV to 0.551~keV corresponds to shock jumps from 246~km~s$^{-1}$ to 676 km~s$^{-1}$. These are
well below the terminal wind speed of 1885~km~s$^{-1}$ (=~2.3~keV), so they seem physically reasonable.

Out of the three $EM$ values the most uncertain is the one corresponding to the lowest temperature 
component ($T$=0.07~keV). This
component is determined by the \ion{C}{\sc vi} and \ion{N}{\sc vi} features. 
The most robust of the three components is the one at 0.20~keV because it involves the largest number of
emission lines in the spectrum. 
The $EM$ values for $\zeta$~Ori decrease with T
and approximately follow the relation $EM \propto T^{-0.8}$.

The total $EM$ over the three temperature bins is 
1.35$\times$ 10$^{55}~$cm$^{-3}$. It is interesting to compare these values to the "wind emission measure" $EM_w$
given by Cassinelli et al. (1981). Based on
their formula we obtain for $\zeta$~Ori $EM_w= 2.44 \times 10^{58}~cm^{-3}$. This shows that only a small fraction of the 
wind is contributing to the X-ray spectrum. 

The emission lines observed in $\zeta$~Ori's X-ray spectrum clearly 
exhibit Doppler broadening of roughly 1000 $\pm$ 100~$\kms$.  
Therefore, the lines generated in the model spectrum need to be 
broadened in excess of instrumental effects to allow a reasonable 
comparison with the data.  For this purpose, we used the 
micro-turbulent velocity parameter ($v_{mic}$) available in SPEX.   
In using this parameter, we do not mean to imply that the emission 
line broadening is microturbulent in nature, rather, it is just a 
convenient way to parameterize the line broadening due to the 
outflow of the X-ray emitting regions.  The $v_{mic}$ parameter is 
related to the FWHM measure of the line broadening through the relation 
$FWHM/E_{obs} = 2\sqrt{\ln 2}v_{mic}/c$. Here $E_{obs}$ is the energy of the emission line,
i.e. $12.39842/\lambda_{obs}$.
A characteristic value used in the fitting 
was $FWHM/E_{obs} = 5.5 \pm  0.5 \times 10^{-3}.$

To achieve the best fit possible, the overall Doppler shift 
parameter $z_{red}$ was used.  The best fits were achieved 
using an overall blueshift of roughly 100~\kms. 
To interpret this number, 
it is interesting to note that some small blueshift of the spectral line 
centroids is to be expected due to the attenuation of the X-rays emitted 
from shocks moving away from the observer on the far side of the wind.
Nonetheless, the fact that 
this shift is roughly of the order of systematic uncertainties in the 
wavelength calibration (Pollock 2007b) means this result should be 
viewed with some caution.   The interpretation of individual emission 
line shapes is considered in more detail in Sect.~3.

\subsubsection{Abundances}
The elemental abundances are relative to solar photospheric values
from optical studies (Anders \& Grevesse, 1989) except for Fe, for which we
use log\,$A_{\rm Fe}$\,=\,7.50\footnote{Here log\,$A_{\rm Fe}$ is the
logarithm of the Fe-abundance relative to log\,$A_{\rm H}$\,=\,12.0.} (see
Grevesse \& Sauval 1998 and 1999) instead of 7.67 (Anders \& Grevesse~1989).

The abundances have been determined through a fit to the total spectrum,
allowing them to freely vary along with the emission measures and temperatures. 
When interpreting these abundance measurements, two issues should be
kept in consideration:  First, the abundances are
strongly anti-correlated with the emission measures (especially in cases involving a weak continuum),
with the result that 
the products of emission measure and abundance ($EM \times A_i$) are quite robust.
Second, the ratios between the abundances ($A_i/A_{\rm O}$ or $A_i/A_{\rm Fe}$) of the various elements are more robust than the absolute
abundances themselves. 
Therefore, the ratios between the abundances are often presented in papers. Here we give both values,
the absolute abundances and the ratio of the elemental abundance over the oxygen abundance averaged for a number of fits.

The carbon abundance is strongly related to the
lowest temperature and the corresponding emission measure. Almost 50\% of the
line flux of this ion is produced by the lowest temperature component. The
obtained abundances are all close to solar photospheric values except for Ne,
Mg, and Si. 
This behaviour is different from coronal plasmas, for which in the quiescent state a First Ionization Potential (FIP) effect is
noticed. This effect implies that elements with low first ionization potential (Mg, Si, Fe) are overabundant in the
corona compared to the photosphere. For active, flaring coronal plasma an inverse FIP effect is often found.  

\subsubsection{X-ray luminosity}
The luminosity is a robust quantity. The X-ray luminosities, given in Table~2, are the model luminosity in the band 0.3~-~10~keV (the energy range
of {\em XMM-Newton}) and in the energy band 0.03~-~0.9~keV, the energy range of the determined emission measure (see Fig.~\ref{DEM}).
Both values correspond to the location of the emitting plasma, i.e.,
they are corrected for absorption by the interstellar medium (ISM) and by the stellar
wind. Our instrumental (0.3~-~10~keV) value of $L_{\rm x} = 1.37 \times 10^{32}\, \ergpers$ results in a value for log$ L_{\rm x}/L_{bol}$ = -6.66(.09)
which is similar to the values obtained using
{\em Einstein} (Chlebowski et al. 1989) and {\em ROSAT} (Bergh\"ofer et al. 1996).

In order to attribute the X-ray emission from this star to shocks embedded in its wind, 
some of the mechanical energy of the wind outflow must be converted into heat energy.    
It is therefore interesting to compare the measured X-ray luminosity to the mechanical 
luminosity of the wind, a quantity which can easily be calculated from a formula given in Howk et al. (2000).
The resulting value of $L_{\rm w} = 1.6 \times 10^{36}~\ergpers$ should be compared 
with the X-ray luminosity of the total emission of $L_{\rm x} = 1.4 \times 10^{33}~\ergpers$. This indicates 
that only about a one thousandth of the mechanical energy available in the wind is radiated 
away as X-rays, a small fraction.

\subsection{DEM modeling}
Apart from a multi-$T$ fitting, the combined {\sc epic-mos}, {\sc epic-pn}, and {\sc rgs} spectrum was also fitted by means of a differential emission 
measure (DEM) modeling (e.g., Kaastra et al. 1996b) using the
regularization module in {\sc spex}. This module constructs an emission measure distribution with the constraint that the first and
second derivatives are smooth and continuous.
We define the $DEM$ by $n_en_{\rm H}dV/d{\rm log}T$ (integrated over
one logarithmic temperature bin, the emission measure = $n_en_{\rm H}V$). 

\begin{figure}[h!]
\hbox{\psfig{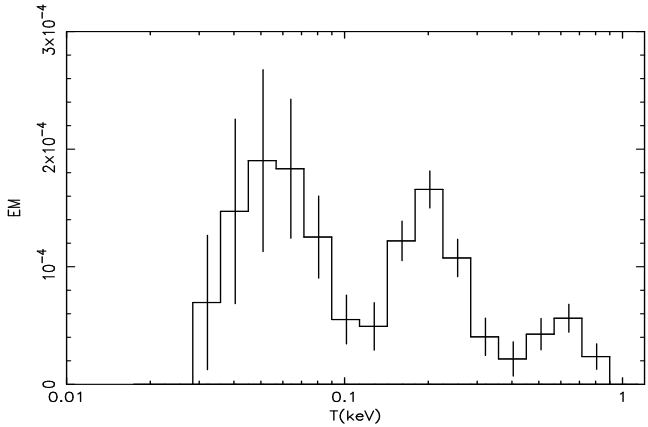}}
      \caption[]{DEM modeling of the {\sc epic-mos} and {\sc rgs} spectra of $\zeta$~Ori. The emission measure
      $EM$ per logarithmic temperature bin is in units of 10$^{58}$~cm$^{-3}$.}  
         \label{DEM}
   \end{figure}

The result for the emission measure distribution $EM$
is shown in Fig.~\ref{DEM}.
The temperature covers the range from 0.03~keV to 0.9~keV, i.e., 
from 0.3 to 10~MK, with three temperature peaks around 0.06~keV, 0.2~keV, and 0.6~keV. These values 
are consistent with the results of the 3-$T$ fittings. Just as in the case of the 3-$T$ fitting the DEM-modeling is
most robust for the range around 0.2~keV, while the error bars are larger for the low temperature regime. 

\begin{table*}
\caption[]{Measured line fluxes at Earth of $\zeta$~Orionis.
Statistical 1$\sigma$ errors in the last digits in parentheses. }
\label{lineflux}
\begin{center}
\begin{tabular}{|l@{\ }|c@{\ }c@{\ }c@{\ }c@{\ }r@{\ }c@{\ }c@{\ }|}    
\hline \hline
ion     &$\lambda_0^a$& $\lambda_{\rm obs}^a$ &${\Delta \lambda}^b$&${\Delta v}^c$ &flux             
&{\sc fwhm}$^d$&{{\sc hwhm}}$^e$  \\
        &[\AA]  &[\AA] &[\AA]   &[km~s$^{-1}$]   & [10$^{-13}$~erg&[10$^{-3}$~keV]    & [km~s$^{-1}$]         \\
        &       &      &        &                &             cm$^{-2}$s$^{-1}$]     &   &                   \\
\hline
EPIC           &       & &      &   &        &    &     \\
Ar\,{\sc xvii} & 3.949 &3.92(.031)  &--   &--      &0.07(.03)&--  & -- \\
S\,{\sc xv}    & 5.039 &5.032(.018) &--   &--      &0.18(.05)&--  & -- \\
Si\,{\sc xiii} & 6.692 &6.684       &--   &--      &0.84(.42)&--  & -- \\
Mg\,{\sc xi}   & 9.170 &9.177(.014) &--   &--      &1.18(.09)&--       & -- \\
RGS            &       &     &  & &  & &   \cr
Si\,{\sc xiii} & 6.692  &6.688(.066) &--   &--      &0.84(.60)&--       & -- \\
Mg\,{\sc xii}  & 8.423  &8.405(.030) &--   & --     & 0.27(.13) & --      & --\\
Mg\,{\sc xi}   & 9.169  &9.169$^f$   &--   & --     & 0.59(.15) &7.5(7.5) &800(800)\\
Mg\,{\sc xi}   & 9.231  &9.231$^f$   &--   & --     & 0.31(.14) &7.5(7.5) &800(800)\\
Mg\,{\sc xi}   & 9.313  &9.313$^f$   &--   & --     & 0.33(.12) &7.5(7.5) &800(800)\\	      
Ne\,{\sc x}    & 10.240 &10.212(.020)&-.028&-820(590)&0.29(.14) &$<$11.0  &$<$1330\\	      
Ne\,{\sc x}    & 12.134 &12.131(.008)&-.003&-70(200)& 1.83(.26) & 4.8(1.9)&700(300)\\
Fe\,{\sc xvii} & 12.264 &12.256(.023)&-.008&-200(600)&0.61(.19) & 4.8(1.9)&700(300)\\
Ne\,{\sc ix}   & 13.447 &13.444(.006)&-.003&-70(130)& 3.35(.39) & 3.8(1.0)&620(160)\\
Ne\,{\sc ix}   & 13.553 &13.548(.009)&-.005&-110(200)&2.26(.32) & 3.8(1.0)&620(160)\\
Ne\,{\sc ix}   & 13.700 &13.699(.017)&-.001&-20(400)& 0.66(.20) & 3.8(1.0)&630(170) \\
Fe\,{\sc xviii}& 14.205 &14.204(.011)&-.001&-20(220)&0.79(.18)  & 3.7(3.0)&640(515) \\    
Fe\,{\sc xvii} & 15.013 &15.012(.004)&-.001&-20(80) & 5.33(.33) & 4.0(0.7)&720(130)  \\
O\,{\sc viii}  & 15.176 &15.176$^f$  & --  & --     & 0.99(.44) & 4.0(0.7)&720(130) \\
Fe\,{\sc xvii} & 15.260 &15.261(.014)&+.001&+20(280)& 2.16(.42) & 4.0(0.7)&720(130) \\
Fe\,{\sc xvii} & 15.449&15.415(.026)&-.034&-700(500)& 0.54(.26)&$<$8.3 &$<$1550 \\    
O\,{\sc viii}  & 16.006 &16.017(.008)&+.011&+210(150)$^h$ & 2.21(.29) & 4.9(1.2)&950(230) \\
Fe\,{\sc xviii}& 16.078 &--&--    & -- & --        &--       & --\\	
Fe\,{\sc xvii} & 16.775 &16.771(.006)&-.004&-70(110) & 2.04(.22) & 2.0(1.1)&410(220) \\
Fe\,{\sc xvii} & 17.051 &17.051$^i$  &--   &--       & 2.84(.52) & 3.7(0.9)&760(190)\\
Fe\,{\sc xvii} & 17.100 &17.100$^i$  &--   &--       & 2.34(.42) & 4.7(1.3)&970(270)\\
O\,{\sc vii}   & 17.396 &17.386(.034)&-.010&-170(600)& 0.62(.23) & --      & --    \\    
O\,{\sc vii}   & 17.768 &17.740(.014)&-.028&-500(250)& 0.64(.12) & 4.0(1.5)&860(320)    \\    
O\,{\sc vii}   & 18.627 &18.617(.009)&-.010&-160(144)& 1.10(.16) & 3.9(0.8)&880(180)\\
O\,{\sc viii}  & 18.969 &18.962(.003)&-.007&-75(50)  &10.78(.36) & 3.5(0.3)&800(70)\\
O\,{\sc vii}   & 21.602 &21.594(.005)&-.008&-110(70) & 8.07(.60) & 2.4(0.3)&630(80)\\
O\,{\sc vii}   & 21.804 &21.775(.007)&-.029&-400(100)& 8.44(.63) & 3.8(0.4)&1000(110)\\
O\,{\sc vii}   & 22.101 &22.088(.009)&-.013&-180(120)& 0.93(.15) & 3.0(0.7)&800(190) \\
N\,{\sc vii}   & 24.781 &24.771(.009)&-.010&-120(110)& 1.44(.14) & 2.5(0.6)&750(180) \\
N\,{\sc vi}    & 24.898 &24.898(.041)&-.000&   0(490)& 0.34(.16) &1.8(0.6) &530(180) \\	
Ca\,{\sc xiii} & 26.719 &26.712(.055)&-.007&-80(640) & 0.13(.08) & $<$2.4  &$<$790  \\    
C\,{\sc vi}    & 26.990 &26.972(.024)&-.018&-200(270)& 0.24(.09) & $<$2.2  &$<$720    \\     
Ar\,{\sc xiv}? & 27.41~~&27.343(.029)&-.067&-700(320)& 0.19(.10) & $<$5.0  &$<$1600   \\     
Ca\,{\sc xii}  & 27.973 &27.955(.027)&-.018&-190(300)& 0.25(.10) & 1.6(1.3)& 540(440) \\     
C\,{\sc vi}    & 28.466 &28.396(.031)&-.070&-700(300)& 0.62(.12) & 2.9(1.5)&1000(520) \\
N\,{\sc vi}    & 28.787 &28.771(.027)&-.016&-170(280)& 0.79(.12) & 3.0(0.6)&1040(210) \\
N\,{\sc vi}    & 29.084 &29.076(.027)&-.009&-90(280) & 0.97(.14) & 3.0(0.6)&1050(210) \\
N\,{\sc vi}    & 29.534 &29.521(.059)&-.013&-130(600)& 0.21(.07) & 3.0(0.6)&1070(210) \\
S\,{\sc xiv}   & 30.441 &30.427(.023)&-.014&-140(230)& 0.16(.08) & $<$1.0  &$<$370    \\     
Si\,{\sc xii}  & 31.016 &30.959(.033)&-.057&-550(320)& 0.17(.07) & $<$2.4  &$<$900   \\     
S\,{\sc xiii}  & 32.22~~&32.226(.030)&--   &--       & 0.22(.08) & $<$1.7  &$<$660    \\     
S\,{\sc xiv}   & 32.554 &32.527(.043)&-.027&-250(400)& 0.28(.09) & $<$2.2  &$<$870    \\     
C\,{\sc vi}    & 33.736 &33.713(.010)&-.023&-200(90) & 2.96(.33) & 1.9(0.3)&770(120) \\
S\,{\sc xiii}  & 35.66~~&35.622(.021)&--   &--       & 0.53(.08) & 1.5(0.5)&650(220)    \\     
S\,{\sc xii}   & 36.398 &36.350(.030)&-.048&-400(250)& 0.40(.10) & 1.3(0.7)&570(310)    \\     

\hline \hline
\end{tabular}
\end{center}
\begin{flushleft}
{
\begin{description}
\item Notes:
\item $a$: $\lambda_0$ is theoretical wavelength from Kelly (1987) and Dere et al. (2001) and
 $\lambda_{\rm obs}$ is observed wavelength with the statistical 1$\sigma$ error in parentheses;
$b$: line wavelength shift; $c$: velocity shift; $d$: line broadening {\sc fwhm}; 
$e$: line half-width at half maximum; $f$: line wavelength fixed to assumed "blue-shifted" value;
$h$: if we take the average wavelength for the 
O\,{\sc viii}, Fe\,{\sc xviii} blend then we obtain a velocity shift of -500~\kms;
$i$: line wavelength fixed to theoretical value. 
\end{description}
}
\end{flushleft}
\end{table*}

\section{Emission line fluxes}
As mentioned in the introduction, the shapes of the X-ray emission 
lines detected by high resolution X-ray spectrometers have been very 
important in determining the nature of the X-ray emission from these stars.
Therefore, apart from the description of the total X-ray spectrum of $\zeta$~Ori by means of a CIE model, the fluxes,
wavelengths, and broadenings of the individual emission lines were measured.
This was done by folding a Gaussian through the response matrix. 

In cases where line blending would interfere with a measurement, we adopted the 
following procedure to isolate the contribution of the line being measured:  
The parameters derived in the multi-T fit (Sect.~2.1.1.) were used to generate a 
model spectrum which included the contributions of all relevant ions \textit{except} 
the ionization state of the line being measured.  A single Gaussian (whose parameters 
were free to vary in the fit) was then added on top of the computed model spectrum 
to measure the individual contribution of the line being measured.
 
In Table~3 the wavelengths, the measured fluxes at Earth (not corrected for interstellar absorption), and the widths of a number of prominent lines
are collected. Within brackets the 1$\sigma$ errors are given.
The rest wavelengths ($\lambda_0$) are taken from Kelly (1987) and Dere et al. (2001),

\subsection{Analysis of emission line profiles}
The lines in the  {\sc rgs} spectrum of $\zeta$~Ori are symmetric, broadened and slightly blue-shifted.
After correcting for instrumental broadening, FWHM is $\approx$ 1500 km~s$^{-1}$ (cf. Table~3). For the dominant, unblended lines of  
\Cvi (33.736 \AA), \Nvii (24.781 \AA), \Oviii (18.969 \AA), \Fexvii (15.013 \AA), and \Nex (12.134 \AA) we obtain a FWHM
 of 1540(240)~\kms, 1500(360)~\kms, 1600(140)~\kms, 1440(260)~\kms, and 1400(600)~\kms, respectively.
Our results confirm the observations done 
with the {\em Chandra}-{\sc hetgs} (Waldron \& Cassinelli 2001, Miller 2002, Miller et al. 2002). 
The line broadenings of the individual line measurements are consistent with the relative broadening $FWHM/E_{obs}$ of 5.5 (in units 0.001),
determined by means of the $v_{\rm mic}$ value of the multi-temperature fit in Sect.~2.1.1. 
For the same lines as mentioned above the values of $FWHM/E_{obs}$ are 5.2(0.8), 5.0(1.2), 5.4(0.5),
4.8(0.8), and 4.7(1.9), respectively. This broadening is consistent with a picture of outwardly moving shocks.

Most wavelengths of the lines in Table~3 show a negative deviation, corresponding to $\approx$
-90(60)~\kms, which is based on the same five lines as used above with individual deviations of -200(90)~\kms,
-120(110)~\kms, -75(50)~\kms, -20(80)~\kms, and -70(200)~\kms. There is some tendency that the features of the
lower ionized (cooler) ions are more blue-shifted. 
However, due to the large errors, the evidence for blue-shifted line profiles is not very strong. 

The possible effects of resonance line scattering opacity on the shapes of 
hot star X-ray emission lines have been discussed in Leutenegger et al. (2007) and
Ignace \& Gayley (2002).
We use the measured line fluxes of \ion{Fe}{\sc xvii} to estimate the optical thickness of 
the shock sources of $\zeta$~Ori. In an optically thin plasma, the line
flux ratio of the \ion{Fe}{\sc xvii} lines at 15.013 \AA\ and 15.260 \AA\ would be expected to be 3.5(1.0) from the MEKAL atomic data, 
while the measured ratio for our $\zeta$~Ori data is 2.5(0.5).
Thus we can conclude that there is no severe resonance line scattering in the \ion{Fe}{\sc xvii} line at 15.013~\AA\ and that the 
individual line source regions are optically thin, even though the entire wind is thick to X-rays at these wavelengths. 

The measured line fluxes of the He-like lines of \ion{N}{\sc vi}, \ion{O}{\sc vii}, \ion{Ne}{\sc ix}, and \ion{Mg}{\sc xi} are used in the next
section.

It is likely the weak features in the {\sc epic-mos} spectrum near 3.9~\AA\ and 5.0~\AA\ are due to \ion{Ar}{\sc xvii}
and \ion{S}{\sc xv}, respectively.
The optimal formation temperatures of these ions (22~MK or 1.9~keV, and 16~MK or 1.4~keV, respectively) are much higher than any of the temperatures found in our 
three-temperature and DEM temperature modeling in Sect.~2. The presence of emission lines from these ions indicates that they are formed in a 
high energy tail in the emission measure distribution of this star and in the wings of the line profiles.

\begin{figure*}
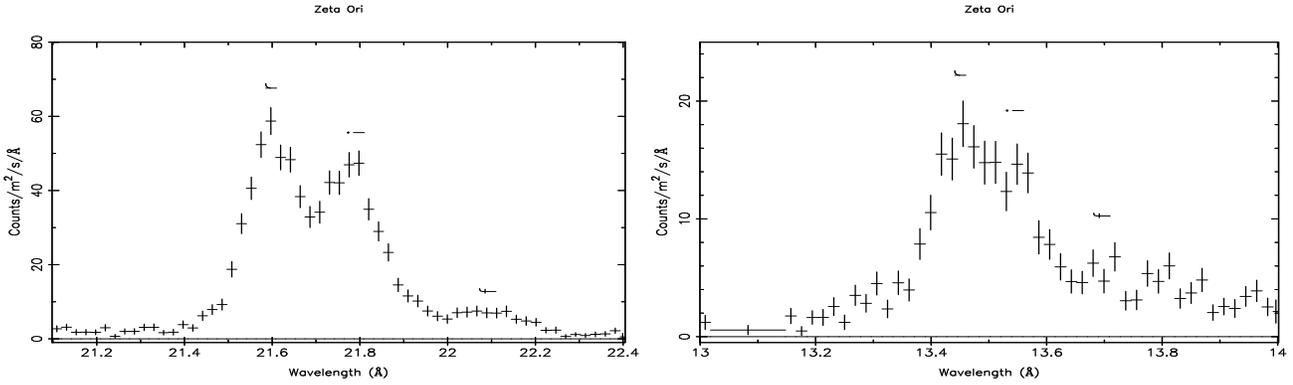

\vbox{
\hbox{
\psfig{figure=Raassenfig3a.eps,angle=-90,height=5cm,width=8.5cm}
\psfig{figure=Raassenfig3b.eps,angle=-90,height=5cm,width=8.5cm}
}
}
 \caption[]{The panels show the He-like \ion{O}{\sc vii} and \ion{Ne}{\sc ix} line triplets consisting of a 
       resonance line (r), an intercombination line (i), and a forbidden line (f) for the O9.5Ib star $\zeta$~Ori.\\
      In both panels, the forbidden line nearly disappears due to radiative depopulation of the
      upper level in the strong radiation field relatively close to the surface of the star.      
              }      
         \label{fir}
   \end{figure*}

\subsection{He-like line diagnostics}
Gabriel \& Jordan (1969) were the first to demonstrate that forbidden
($f$), intercombination ($i$) and resonance ($r$) lines of the He-like
``complexes" provide useful diagnostics for X-ray emitting plasmas. The most
recent studies including the effects of dielectronic satellite lines and a
radiation field have been done by Porquet et al. (2001). The ratio $f/i$ is
dependent on electron density because of the depopulation of the upper level
of the $f$ line in favour of the upper level of the $i$ line at increasing density. 
This effect can be used to derive
electron densities in circumstances where radiation fields are relatively
weak, such as in cool stars. However, in OB stars, where the radiation fields
are much stronger, the depopulation of the upper level of the $f$ line 
in favour of the upper level of the $i$ line
occurs by radiative absorption (e.g., Blumenthal et al. 1972, Porquet
et al. 2001). The $f/i$ ratio no longer indicates the density, but instead
provides information on the mean intensity of the radiation field, hence the
radial distance $R$ of the X-ray source from the star.
This is illustrated in Fig.~\ref{fir} for the \Neix and \Ovii triplets.
It is seen that the forbidden line is nearly suppressed for both ions.

  The fact that the UV radiation field is the dominant effect for $\zeta$~Ori is
confirmed by detailed calculations.  
Using the formalism developed by Blumenthal et al. (1972), we calculate the radial dependence of $R$
in the envelope of $\zeta$~Ori on the basis of a Planck curve (Ness et al. 2001) of 31500~K as well as on the basis 
of a photospheric model 
of Vacca et al. (1996) with $T_{\rm eff}$=32000~K and log $g$=3.23 and applying a UV flux model   
OSTAR2002 of Lanz \& Hubeny (2003) with $T_{\rm eff}$=32500~K, log $g$=3.25, and solar metallicity.
These parameters closely correspond to those given in the appendix.
The radial-dependent wind density is derived from the stellar and wind parameters
of Lamers \& Leitherer (1993) using a standard velocity law that is modified below 1.02$R_*$ (where $R_*$ is 
the stellar radius) to provide a smooth density transition to the photospheric structure. 
The mean intensity of the UV radiation is large near the surface of the star and it decreases outwards by
dilution factor  
$W(r)=\frac{1}{2}~\left[1-\left(1-\left(\frac{R_{*}}{r}\right)^2\right)^{1/2}\right]$.
As a result of the radial dependence of the radiation field, the observed $f/i$ ratio can be used to derive
the radial location of the He-like ions that are producing the observed $fir$ lines. 
For the four He-like ions which allow reliable measurements of their
$f/i$ line ratio (\ion{N}{\sc vi}, \ion{O}{\sc vii}, \ion{Ne}{\sc ix}, and \ion{Mg}{\sc xi}) these parameters indicate that
$\zeta$~Ori's radiation field will suppress the forbidden lines by radiative de-excitation of the
upper level of $f$ to much greater
radii than would be possible with collisions (see, e.g., Waldron \& Cassinelli 2001). 
Thus the expected $f/i$
ratios as functions of radii for these ions are entirely
controlled by the strength of $\zeta$~Ori's UV radiation field.
 
\begin{figure}
\center
{\psfig{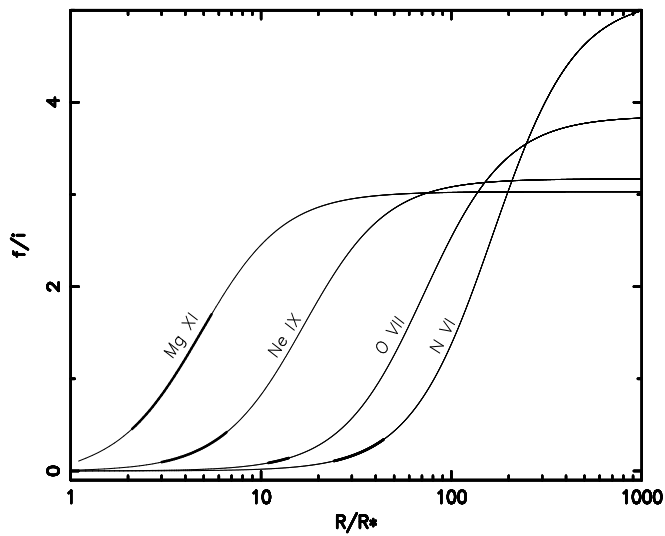}}
\caption[]
{The curves show the expected $f/i$ line ratio as a
function of formation radius $R$ for the labeled He-like ions.
The thickened portion of each line gives the radial range of the X-ray
emission implied by the measured $f/i$ line ratios.
\label{fi-ratio}}
\end{figure}

The predicted $R$ dependencies of $f/i$ are shown in Fig.~\ref{fi-ratio}.
The $f/i$ ratios derived from the
{\sc rgs} spectra for \ion{N}{\sc vi}, \ion{O}{\sc vii}, \ion{Ne}{\sc ix}, and \ion{Mg}{\sc xi} are:
0.22$\pm$0.12, 0.114$\pm$0.027, 0.26$\pm$0.16, and 1.08$\pm$0.61, respectively (cf. Table~3), agreeing with values
for \ion{Ne}{\sc ix} and \ion{Mg}{\sc xi} given by Oskinova et al. (2006). 
These ratios correspond to average distances from the stellar surface $R$ of the ion formation
of 34$\pm$10$R_*$, 12.5$\pm$1.5$R_*$, 4.8$\pm$1.8$R_*$, and 3.9$\pm$1.7$R_*$, respectively. For the last three ions
this is in agreement with values established by Waldron \& Cassinelli (2001), based on HETGS observations. 
They determined values for formation radii of $\la$12$R_*$ for \ion{O}{\sc vii}, of  $\la$6$R_*$ for \ion{Ne}{\sc ix}, and
3--5$R_*$ for \ion{Mg}{\sc xi}. 

The formation distances of He-like ions increase with decreasing ionization stage, but they increase also with increasing wavelength ($\lambda$) of the
line triplets. The formation distance is proportional to some power of the wavelength of the intercombination line ($\lambda^p$, with $p\approx 2.5-4.5$).
 Waldron \& Cassinelli (2001) pointed out that the derived radii ($R_{fir}$), at which 
the He-like lines are observed, are strongly
correlated with the radii $R_\lambda$ at which the continuum adjacent to the line has an optical depth unity.
Since the opacity varies roughly as $\lambda^3$ this correlation would mean that we are observing the He-like lines 
from the deepest possible part of the wind from which X-rays could escape to the observer. 

Leutenegger \ea (2006) give estimates for minimum radii of formation of
the hot plasma for O-stars. Their
values for minimum radii are about 1.5$R_*$, based on the ions of higher stages of ionization
(\ion{S}{\sc xv}, \ion{Si}{\sc xiii}, and \ion{Mg}{\sc xi}). 
The reader should keep in mind that our measurements of the radii of 
line formation are not directly comparable to theirs.
Our values are average values with a lower and higher limit due to the
uncertainty in the measured line fluxes and do not imply a minimum or maximum distance of ion 
formation.

Finally, we used the resonance line, the forbidden line, and the
intercombination line of the He-like lines as a temperature diagnostic
for the individual regions responsible for the fir emission of each
ion.  As a single-temperature measurement method for specific regions
in the wind, this method gives independent, complementary information
to the global temperature fitting carried out in Sect.~2.
Using the results of Porquet et al. (2001) we extract from the measured ($i+f$)/$r$ ratios (cf. Table~3) 
for \Nvi, \Ovii, and \Neix electron temperatures of about 0.05, 0.16, and 0.24~keV,
respectively. These three temperatures are in the range of the lower two of the three
temperatures found with multi-temperature fitting in Sect.~2.1.1.

\section{Discussion of results and conclusion}
The O supergiant $\zeta$~Orionis (O9.7 Ib) has a line rich X-ray spectrum. 
The spectral lines are symmetric and broadened.
The FWHM corresponds to 1500~km~s$^{-1}$;
less than the terminal speed of 1885 km~s$^{-1}$ derived by Lamers et al. (1999). 
The velocity widths are consistent 
with a picture of outwardly propagating 
shocks that are distributed throughout the wind and that are 
produced by the instabilities associated with a line-driven wind.
The best fits require a small blue-shift. It should be noted, however,
that this blue-shift is of the order of the calibration uncertainties
of the instrument.

The continuum and
line features are very well described by a 3-T collisional ionization equilibrium model (CIE) for optically thin plasma,
using SPEX in combination with MEKAL. The determined temperatures range from 0.07 to 0.55~keV (0.85 - 6.4~MK), corresponding to
shock jumps from 246~km~s$^{-1}$ to 676~km~s$^{-1}$. The total X-ray luminosity of the star
corresponds to an emission measure that is far below the emission measure
of the entire wind (Cassinelli et al. 1981, Cassinelli \& Swank 1983), 
indicating that only a small fraction of the wind is participating in the X-ray emission.

We used the values of the forbidden to intercombination line ratio of 
He-like ions to derive characteristic radii of line formation.
For \ion{Mg}{\sc xi}, \ion{Ne}{\sc ix}, 
\ion{O}{\sc vii}, and \ion{N}{\sc vi}, this resulted in values of 4, 5, 12.5, and 34 stellar radii, respectively.
 
The hotter and higher ionized ions are on average 
formed deeper in the wind, closer to the stellar surface. 
This is in agreement with values established by Waldron \& Cassinelli (2001), based on HETGS observations. 
Applying the standard beta velocity law  $v(r)=v_{\infty}\left( 1-\frac{R_*}{r}\right)^{\beta},$ with
$\beta$ = 0.7(.1) (Groenewegen \& Lamers 1989) we obtain at a distance of 4$R_*$ (where the hotter plasma of \Mgxi is
located) $v(r)=0.82\times v_{\infty}=1540~\kms$. This velocity is far higher than the velocity jump necessary to
produce the high temperatures.

Waldron and Cassinelli (2001) found that line diagnostics for \ion{Si}{\sc xiii} indicate that this line emission forms 
very close to the stellar surface, where the velocity is too small to produce the shock jump required 
for the observed ionization level.
The analysis of 17 O stars by 
Waldron \& Cassinelli (2007) has led to a strengthening of this conclusion. 
However, due to the lower resolution of the {\sc rgs} relative to that of the {\sc hetgs} on {\em Chandra} and 
the weakness of the relevant lines in our spectrum, we cannot investigate the formation radius of \Sixiii. 
These limitations therefore preclude us from confirming or denying the presence of this 
effect in the X-ray emission of $\zeta$~Ori. 

We conclude that our results are consistent with a model
of shock fragments that are embedded in the wind and are expanding outward with the wind. The \Fexvii line ratios indicate that the
individual source regions associated with the shock fragments are optically thin. 
The lack of variability of the X-rays over the years 
implies a large number of sources embedded in the wind. Ions are found to be formed and emitting both from far out and deep in the wind.
The symmetry of the line profiles and the slight blue-shift are consistent with the ``porosity model" 
(Feldmeier et al. 2003, Oskinova et al. 2006, Owocki \& Cohen 2006 and Fig.~4 therein), though we find a wider range in 
line formation radii than the $\approx 1.5~R_*$ used in their scenario.

\begin{acknowledgements}
This work is based on observations obtained with {\em XMM-Newton}, an ESA science 
mission with instruments and contributions directly funded by 
ESA Member States and the USA (NASA).
The SRON Netherlands Institute for Space Research is supported financially by NWO.
JPC and NAM  acknowledge support from NASA grants NAG5 5-9226 and GO2-3028, respectively.
 \end{acknowledgements}

\newpage
\appendix
\section{}

% TABLE 1 %%%%%
\begin{table*}[h!]
\caption{Stellar parameters of $\zeta$\,Ori}
\small
\vspace*{-3mm}
\begin{center}
\begin{tabular}{l | c l | c l | c l | c l}
\hline\hline
                                                    &                         &      &                   &      &                    &       &                    &       \\[-2mm]
parameter                                           & $\zeta$\,Ori~Aa         &      & $\zeta$\,Ori~Ab   &      & $\zeta$\,Ori~B     &       & $\zeta$\,Ori~C     &       \\
                                                    &                         & ref. &                   & ref. &                    & ref.  &                    & ref.  \\
                                                    &                         &      &                   &      &                    &       &                    &       \\[-2mm]
\hline
                                                    &                         &      &                   &      &                    &       &                    &       \\[-2mm]
HR                                                  & 1948                    &      &                   &      & 1949               &       &                    &       \\
HD                                                  & 37742                   &      &                   &      & 37743              &       &                    &       \\
ADS                                                 & 4263\,A                 &      &                   &      & 4263\,B            &       & 4263\,C            &       \\[-2mm]
                                                    &                         &      &                   &      &                    &       &                    &       \\
\hline
                                                    & \multicolumn{7}{c}{}                                                                                        &       \\[-2mm]
$d$              (kpc)                              & \multicolumn{7}{c}{0.25$^{+0.06}_{-0.04}$\,$^a$}                                                            & ESA97 \\
$E_{B-V}$        (mag)                              & \multicolumn{7}{c}{0.06}                                                                                    & VB89  \\
$A_V$            (mag)                              & \multicolumn{7}{c}{0.19}                                                                                    &       \\
log\,$N_{\rm H}$ (cm$^2$) (ISM)                     & \multicolumn{7}{c}{20.34}                                                                                   & FP97  \\
                                                    & \multicolumn{7}{c}{}                                                                                        &       \\[-2mm]
\hline
                                                    &                         &      &                   &      &                    &       &                    &       \\[-2mm]
separation       (arcsec)                           &                         &      & 0.045             & HW00 & 2.42               & MG98  & 57.6               & MG98  \\
$P$              (yr)                               &                         &      &                   &      & 1509               & MG98  &                    &       \\
spectral type                                       & O9.7\,Ib                & WF90 & late O\,V         & HW00 & B2\,III            &       &                    &       \\
$\theta_{\rm D}$ (mas)                              & 0.48\,$\pm$\,0.04       & HD74 &                   &      &                    &       &                    &       \\
$V$              (mag)                              & 2.03                    & VB89 & 4                 & HW00 &  4.2               & HW00  &  9                 & HW00  \\
$B-V$            (mag)                              & $-$0.21                 & VB89 &                   &      &                    &       &                    &       \\
$(B-V)_{\rm o}$  (mag)                              & $-$0.27                 & VB89 &                   &      &                    &       &                    &       \\
$v_{\rm rad}$    (km\,s$^{-1}$)                     & $+$25\,$\pm$\,5         & BG78 &                   &      &                    &       &                    &       \\
$v$\,sin\,$i$    (km\,s$^{-1}$)                     & 123\,$\pm$\,3           & Pe96 &                   &      &                    &       &                    &       \\
$v_{\infty}$     (km\,s$^{-1}$)                     & 1885                    & KH96 &                   &      &                    &       &                    &       \\
$T_*$            (kK)                               & 31.5\,$\pm$\,1          & FP97 &                   &      &                    &       &                    &       \\
$M$              (M$_\odot$)                        & 34                      & VB89 &                   &      &                    &       &                    &       \\
log\,$g$         (cgs)                              & 3.2\,$\pm$\,0.1         & VB89 &                   &      &                    &       &                    &       \\
$BC$             (mag)                              & $-3.16$                 & VB89 &                   &      &                    &       &                    &       \\
log $L$/L$_\odot$                                   & 5.2\,$^b$               & FP97 &                   &      &                    &       &                    &       \\
$\dot{M}$ (10$^{-6}$\,M$_\odot\,$yr$^{-1}$)         & 1.4\,$^b$               & FP97 &                   &      &                    &       &                    &       \\
$Y$\,=~$n_{\rm He}$/$n_{\rm H}$                     & 0.1                     & VB89 &                   &      &                    &       &                    &       \\
$A_{\rm N}$                                         & normal                  & BL82 &                   &      &                    &       &                    &       \\
$T_{\rm x}^{\rm {\sc Einstein-ipc}}$ (keV)                & 0.46\,$\pm$\,0.08       & CH89 &                   &      &                    &       &                    &       \\
log\,$L_{\rm x}^{\rm {\sc Einstein-ipc}}$ (erg\,s$^{-1}$) & 32.1\,$\pm$\,0.24$^{b}$ & CH89 &                   &      &                    &       &                    &       \\
$T_{\rm x}^{\rm {\sc ROSAT}}$ (keV)                       & 0.22                    & BS96 &                   &      &                    &       &                    &       \\
log\,$L_{\rm x}^{\rm {\sc Rosat}}$ (erg\,s$^{-1}$)        & 32.0\,$^{b,c}$          & BS96 &                   &      &                    &       &                    &       \\[-2mm]
                                                    &                         &      &                   &      &                    &       &                    &       \\
\hline\hline
\end{tabular}
\end{center}

\footnotesize
{\it Notes:}\\
$a$: The distance to $\zeta$\,Ori is subject to debate.  The three bright
Belt stars in Orion, $\delta$\,Ori (O9.5\,II+B0.5\,III), $\epsilon$\,Ori
(B0\,Ia), and $\zeta$\,Ori, form the high-luminosity end of the Ori\,OB1b
subgroup, which has an estimated age between 7\,Myr (Blaauw 1991) and
1.7\,Myr (Brown et al.  1994).  The distance to Ori\,OB1b had been
estimated to be 360\,$\pm$\,70\,pc based on photometry (Brown et al.
1994), or 473\,$\pm$\,33\,pc based on {\sl Hipparcos} parallax and proper
motion measurements (de Zeeuw et al.  1999).  We adopt here the direct {\sl
Hipparcos} parallax measurement of $\zeta$\,Ori, corresponding to the
distance as given above.                                  \\
$b$: Paramater scaled  to $d$\,=\,0.25\,kpc.              \\
$c$: Bergh\"ofer \& Schmitt (1994) report that $\zeta$\,Ori showed no
evidence for variability over a three year time span, except for a two day
period in September 23-25, 1992, when an increase of $\sim$\,15\% in the
X-ray count rate was measured.

\smallskip

{\it References:} \\
BG78: Bohannan \& Garmany 1978;
BL82: Bisiacchi et al. 1982;
BS96: Bergh\"ofer et al. 1996;
CH89: Chlebowski et al. 1989;
ESA97: ESA 1997;
FP97: Feldmeier et al. 1997;
HD74: Hanbury-Brown et al. 1974;
HW00: Hummel et al. 2000;
KH96: Kaper et al. 1996;
MG98: Mason et al. 1998;
Pe96: Penny 1996;
VB89: Voels et al. 1989;
WF90: Walborn \& Fitzpatrick 1990.

\label{tab:param}
\end{table*}

References:

- Bergh\"ofer, T.W., Schmitt, J.H.M.M., 1994, A\&A 290, 435                 \\
- Bergh\"ofer, T.W., Schmitt, J.H.M.M., Cassinelli, J.P. 1996,
  A\&AS 118, 481 (Erratum: A\&A 121, 212)                             \\
- Bisiacchi, G.F., L\'opez, J.A, \& Firmani, C. 1982, A\&A 107, 252   \\
- Blaauw, A. 1991, in: C.J. Lada \& N.D. Kylafis (eds.), {\it The
  Physics of Star Formation and Early Stellar Evolution},
  NATO ASI Ser. C 342 (Dordrecht: Kluwer), p.\,125                    \\
- Bohannan, B., \& Garmany, C.D. 1978, ApJ 223, 908                   \\
- Brown, A.G.A., de Geus, E.J., \& de Zeeuw, P.T. 1994, A\&A 289, 101 \\
- Chlebowski, T., Harnden, F.R., \& Sciortino, S. 1989, ApJ 341, 427  \\
- ESA 1997, {\it The {\sl Hipparcos} Catalogue}, ESA SP-1200          \\
- Feldmeier, A., Puls, J., \& Pauldrach, A.W.A. 1997, A\&A 322, 878   \\
- Hanbury-Brown, R., Davis, J., \& Allen, L.R. 1974, MNRAS 167, 121   \\
- Hummel, C.A., White, N.M., Elias, N.M., Hajian, A.R., \& Nordgren, T.E. 2000, ApJL 540, L91 \\
- Kaper, L., Henrichs, H.F., Nichols, J.S., Snoek, L.C., Volten, H., 
  \& Zwarthoed, G.A.A. 1996, A\&AS 116, 257  \\
- Mason, B.D., Gies, D.R., Hartkopf, W.I., Bagnuolo, W.G., den
  Brummelaar, T., \& McAlister, H.A. 1998, AJ 115, 821                \\
- Penny, L.R. 1996, ApJ 463, 737                                      \\
- Voels, S.A., Bohannan, B., Abbott, D.C., \& Hummer, D.G. 1989, ApJ 340, 1073 \\
- Walborn, N.R., \& Fitzpatrick, E.L. 1990, PASP 102, 379             \\
- de Zeeuw, P.T., Hoogerwerf, R., de Bruijne, J.H.J., Brown, A.G.A.,
  \& Blaauw, A. 1999, AJ 117, 354                                     \\

\end{document}